\documentclass[a4paper,fleqn,usenatbib]{mnras}

\usepackage{newtxtext,newtxmath}

\usepackage[T1]{fontenc}
\usepackage{ae,aecompl}

\usepackage{graphicx}	
\usepackage{amsmath}	
\usepackage{amssymb}	
\usepackage{xspace}
\hypersetup{final=true}

\newcommand{\Siggas}{\ensuremath{\Sigma_\mathrm{g}}\xspace}
\newcommand{\rhos}{\ensuremath{\rho_\text{s}}\xspace}
\newcommand{\uf}{\ensuremath{u_\text{f}}\xspace}

\newcommand{\Sigdust}{\ensuremath{\Sigma_\mathrm{d}}\xspace}

\title[The disc flux-radius correlation]{On the millimetre continuum flux-radius correlation of proto-planetary discs}

\author[G. P. Rosotti et al.]{Giovanni P. Rosotti,$^{1,2}$\thanks{E-mail: rosotti@ast.cam.ac.uk}, Richard A. Booth$^{1}$, Marco Tazzari$^{1}$, Cathie Clarke$^{1}$, \newauthor Giuseppe Lodato$^{3}$, Leonardo Testi$^{4}$ 
\\
$^{1}$Institute of Astronomy, Madingley Road, Cambridge, CB3 0HA, UK\\
$^{2}$Leiden Observatory, Leiden University, P.O.~Box 9531, NL-2300~RA Leiden, the Netherlands\\
$^{3}$Universit\`a degli Studi di Milano, Via Giovanni Celoria 16, I-20133 Milano, Italy\\
$^{4}$European Southern Observatory, Karl-Schwarzschild-Str 2, D-85748 Garching, Germany
}

\date{Accepted 2019 April 26. Received 2019 March 15; in original form 2018 December 10}

\pubyear{2019}

\begin{document}
\label{firstpage}
\pagerange{\pageref{firstpage}--\pageref{lastpage}}
\maketitle

\begin{abstract}
A correlation between proto-planetary disc radii and sub-mm fluxes has been recently reported. In this Letter we show that the correlation is a sensitive probe of grain growth processes. Using models of grain growth and drift, we have shown in a companion paper that the observed disc radii trace where the dust grains are large enough to have a significant sub-mm opacity. We show that the observed correlation emerges naturally if the maximum grain size is set by radial drift, implying relatively low values of the viscous $\alpha$ parameter $ \lesssim 0.001$. In this case the relation has an almost universal normalisation, while if the grain size is set by fragmentation the flux at a given radius depends on the dust-to-gas ratio. We highlight two observational consequences of the fact that radial drift limits the grain size. The first is that the dust masses measured from the sub-mm could be overestimated by a factor of a few. The second is that the correlation should be present also at longer wavelengths (e.g. 3mm), with a normalisation factor that scales as the square of the observing frequency as in the optically thick case.
\end{abstract}




\section{Introduction}
Planet formation takes place in proto-planetary discs, which provide the building blocks (gas and solids) for assembling the numerous planetary systems observed around main sequence stars. Characterising proto-planetary discs is therefore of fundamental importance for understanding planet formation.

Thanks to the Atacama Large Millimetre Array (ALMA), it is now possible \citep[e.g.,][]{AnsdellLupusI,Pascucci2016} to build statistical inventories of disc properties in various star forming regions. The raw values provided by the surveys are of extreme importance for planet formation models (for example, the solid mass available to turn into planets, see discussion in \citealt{MordasiniReview}). More broadly, the usefulness of these surveys is to test and inform theories of disc evolution \citep[e.g.,][]{Manara2016,Rosotti2017,Lodato2017,Mulders2017}.

The processes controlling how the dust moves and coagulates in the disc (see \citealt{TestiPPVI} for a review) are certainly among the biggest unknowns in disc evolution. Dust growth is obviously a necessary step to form planets, but models of dust coagulation encounter numerous ``barriers'' \citep[e.g.,][]{Brauer2008,Zsom2010,Okuzumi2011,Booth2018} that inhibit growth beyond a certain size. As summarised by \citet{Birnstiel2012}, the two most prominent barriers are those imposed by fragmentation and drift. The former is a consequence of the motions induced by turbulence, which causes the grains to collide and shatter \citep{Voelk1980}, and the latter is a consequence of the fast radial motion of dust grains \citep{Weidenschilling1977}.

Recently \citet{Tripathi} reported the discovery of a quadratic correlation between the disc flux and radius at 850 $\mu$m using results from a previous generation telescope, the Sub-Millimiter Array (SMA). \citet{Andrews2018} confirmed that the correlation is also present using ALMA data in the Lupus star forming region. The same correlation is tentatively present also in the Upper Sco region \citep{Barenfeld2017}, although the method of analysis was different in this case. Pending a homogeneous analysis, in this Letter we will work under the assumption that the correlation is universal.

One possible interpretation of the correlation is that proto-planetary discs are (at least marginally) optically thick. In this Letter we show that another interpretation of the correlation is that it is a sensitive probe of grain growth processes. After first dismissing the hypothesis that the correlation is driven by instrumental sensitivity (section \ref{sec:spurious}), we then argue (section \ref{sec:predictions}) that the correlation emerges naturally, with the correct normalisation, if the grain size is limited by radial drift. In contrast, there is no reason for the observed universal correlation if dust growth is limited by fragmentation. We highlight two consequences of the drift dominated regime in section \ref{sec:consequences} and finally draw our conclusions in section \ref{sec:conclusions}.

\section{Is the flux-radius correlation physical?}
\label{sec:spurious}

Before exploring the possible origins of the correlation, it is worth asking if the correlation is genuine. Of particular concern is the fact that radio interferometers are only sensitive above a given threshold in surface brightness. Given that most of the disc flux is in the outer parts of the disc, a finite surface brightness sensitivity (i.e., the disc ``disappears into the noise'') also leads to a quadratic correlation between disc flux and size (see appendix \ref{sec:sensitivity_cut} for an example).

There are two key predictions of such a hypothesis:
\begin{enumerate}
    \item The normalisation of the correlation (i.e., the average surface brightness) should be, apart from a factor of order unity, the surface brightness sensitivity.
    \item Observations with different surface brightness sensitivities should therefore find different normalisations. Gathering observations with different sensitvities should introduce spread.
   
\end{enumerate}

For what concerns the first prediction, the average surface brightness reported by \citet{Tripathi}, 0.2 Jy/arcsec$^2$, is a factor of 20 higher than the median sensitivity of 5.7 mJy/arcsec$^2$. This cannot be reconciled with a factor of order unity.

With regards to the second prediction, \citet{Tripathi} and \citet{Andrews2018} do not find any statistically significant difference in the correlation normalisation. These two works use very different datasets: the first is a heterogeneous collection of SMA data, while the second is a homogeneous ALMA survey. The ALMA data has a surface brightness sensitivity of 2 mJy/arcsec$^2$. Although the difference in surface brightness sensitivities is modest, $\sim 3$, if the correlation was due to sensitivity effects there should be a discernible difference in the normalisation, given the uncertainties quoted in \citet{Andrews2018}. Such a difference is not observed. We should also expect a higher spread in the SMA data, but \citet{Andrews2018} report instead a comparable scatter around the best fit.


Based on these considerations, we dismiss the hypothesis that the correlation is a spurious consequence of the finite surface brightness sensitivity of the observations. Qualitatively, this is also confirmed by visual inspection of the fitted profiles reported by \citet{Tripathi} and by \citet{Andrews2018}: many show a sudden drop outside some radius, lending credence to the fact that the disc does not simply disappear into the noise. 

\section{Predictions from grain growth models}

\label{sec:predictions}

In this section we present 1D models of dust growth and evolution in proto-planetary discs to explore the origin of the observed disc radius--flux correlation. Our models resemble those presented in \citet{Tripathi}, but here we also seek to provide explanations as to why models of grain growth predict correlations between the disc radius and flux. The models are described in considerable detail in \citet{Booth2017} and in a companion paper \citep{Rosotti2018_rdisc}. In short, we solve the viscous evolution equation for the gas, while for the dust we use the simplified treatment of grain growth described in \citet{Birnstiel2012} {assuming a temperature profile $T=40 \ (r/10 \mathrm{au})^{-0.5} \ \mathrm{K}$ \citep[e.g.,][]{Dalessio1998}.} As a post-processing step, we compute the opacity at ALMA wavelengths resulting from the dust properties obtained from the grain growth model and use it to generate synthetic surface brightness profiles at 850 $\mu$m. Following \citet{Tripathi}, we define the disc radius as the radius enclosing 68 per cent of the total disc flux.

\begin{figure}
\includegraphics[width=\columnwidth]{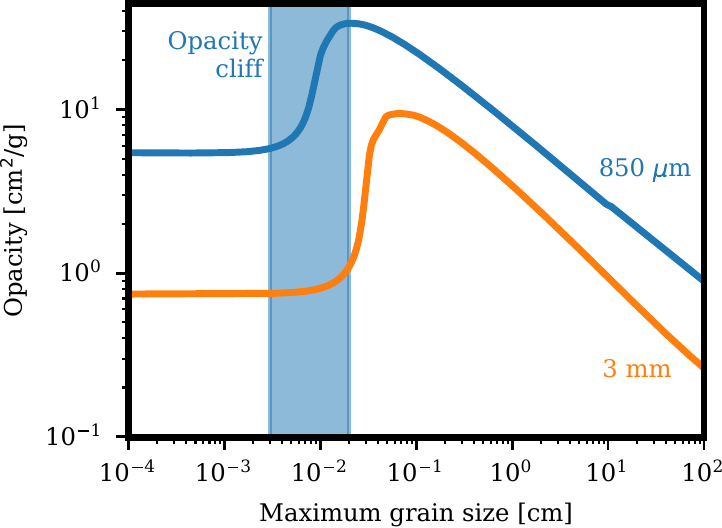}
\caption{The dust opacity at 850 $\mu$m (blue) and at 3 mm (orange) as a function of the maximum grain size. We marked on the figure the ``opacity cliff'' (where the opacity steeply drops by one order of magnitude over a small range of variation in grain size; see text) at a wavelength of 850 $\mu$m.}
\label{fig:opacity}
\end{figure}

There are two aspects of the model that require special attention in light of the following discussion. The first is that in the \citet{Birnstiel2012} model the grain size at each radius is set either by radial drift or by fragmentation. In the former case the maximum radius $a$ of a dust grain is given by 
\begin{equation}
a_\mathrm{drift} = f_\mathrm{d} \, \frac{2 \, \Sigdust}{\pi \, \rhos}\frac{V_\mathrm{k}^2}{c_s^2} \, \gamma^{-1},
\label{eq:a_drift}
\end{equation}
where $f_\mathrm{d}=0.55$ \citep{Birnstiel2012} is an order of unity factor, $\Sigdust$ is the dust surface density, $V_\mathrm{k}$ is the Keplerian velocity for 1 $M_\odot$ star, $\rhos=1$g/cm$^2$ is the dust bulk density, $c_s$ is the gas sound speed, set from the aspect ratio $H/R=0.033 (R/au)^{1/4}$, and $\gamma = | \mathrm{d}\log P / \mathrm{d} \log r |$ is the local power-law slope of the gas pressure. 
In the other regime, the maximum grain size is given by
\begin{equation}
a_\mathrm{frag} = f_\mathrm{f}\, \frac{2}{3 \, \pi} \, \frac{\Siggas}{\rhos \alpha} \, \frac{\uf^2}{c_s^2},
\label{eq:a_frag}
\end{equation}
where $\Siggas$ is the local gas surface density, $f_\mathrm{f}$ is another dimensionless factor (we fix it to 0.37 following \citealt{Birnstiel2012}), $\alpha$ is the \citet{ShakuraSunyaev1973} parametrization of the viscosity (see later) and $\uf$ is the fragmentation velocity, which we set to 10 m/s.

The two limits of \autoref{eq:a_drift} and \autoref{eq:a_frag} have a different dependence on radius; in general the inner parts of the disc are dominated by fragmentation and the outer ones by drift. The crucial parameter in setting the transition radius between the two regimes is the viscosity $\alpha$; highly viscous models are everywhere in the fragmentation dominated regime, while reducing $\alpha$ shifts the transition to lower radii. For practical purposes, models with $\alpha \lesssim 10^{-3}$ are drift dominated down to a few au. Models with $\alpha \sim 0.01$ are initially fragmentation dominated, but become drift dominated after Myr timescales because of the dust surface density reduction.

The second aspect to highlight, because of its observational importance, is the opacity. We compute the opacity as in \citet{2016A&A...588A..53T} following \citet{Natta:2004yu} and \citet{Natta:2007ye}, using the Mie theory for compact spherical grains, assuming a composition of 10\% silicates, 30\% refractory organics, and 60\% water ice. We assume that the grain size distribution is a power-law $n(a)\propto a^{-q}$ with an exponent $q=3.5$, but we do not find significant differences if using $q=3$. In \autoref{fig:opacity} we plot the resulting opacity as a function of the maximum grain size. The most notable feature is the abrupt change in opacity around the characteristic size of 0.2 mm where the maximum 850 $\mu$m opacity is attained. We will refer to this sharp drop (roughly a factor of 10) in opacity as the ``\textit{opacity cliff}''. In the companion paper we show that, with the typical sensitivities of current surveys, the measured disc radii trace the radius where the grains become smaller than their value at the cliff, rather than the physical extent of the disc. Note that the opacity cliff is not present for ``fluffy'' rather than compact grains \citep{Kataoka:2014aa}. Our growth model by construction considers compact grains and therefore we do not consider this possibility further.

\subsection{Model expectations}

\subsubsection{Drift dominated regime}
\label{sec:drift}


Considering in \autoref{eq:a_drift} only the dependence on time or radius:
\begin{equation}
a_\mathrm{drift} \propto   \left(  \frac{H}{R} \right)^{-2} \Sigdust,
\end{equation}
where $H/R$ is the disc aspect ratio. The dust surface density decreases with time due to radial drift and therefore the dust grains become smaller at any given radius. However, a given grain size is always attained at the same surface density (apart from differences due to the radial dependence of the disc aspect ratio). It is of particular interest to consider the grain size $a_\mathrm{cliff}$ corresponding to the maximum opacity (see \autoref{fig:opacity}). In the rest of this section we consider a wavelength of 850 $\mu$m for comparison with the observed correlation, but our theoretical argument holds also at other wavelengths. We elaborate on the consequences of this in section \ref{sec:longer_wavs}. The grains have the critical size at a dust surface density $\Sigma_{d,\mathrm{cliff}}$:
\begin{equation}
\Sigma_{d,\mathrm{cliff}} \propto a_\mathrm{cliff} \left(  \frac{H}{R} \right)^{2} =  a_\mathrm{cliff} R_\mathrm{cliff}^{1/2}, \label{Eqn:SigD}
\end{equation}
where we called $R_\mathrm{cliff}$ the radius where $a = a_\mathrm{cliff}$. Since the flux is dominated by emission at large radii, we can write using the Rayleigh-Jeans approximation and assuming optically thin emission:
\begin{equation}
F_\nu \approx \upi B_\nu(T) \Sigma_d \kappa_\nu R_{\rm cliff}^2 \propto \Sigma_{d,\mathrm{cliff}} R_\mathrm{cliff}^{2} T \propto R_\mathrm{cliff}^{2}, \label{Eqn:FluxR}
\end{equation}
i.e. a quadratic relation between the sub-mm flux and the cliff radius, because the radial dependence of the surface density at the cliff radius is cancelled by that of the temperature.

Note that we not only predict a quadratic correlation, but also in \autoref{eq:a_drift} there are relatively few parameters that can set the normalisation, predicting relatively little scatter.

\begin{figure}
\includegraphics[width=\columnwidth]{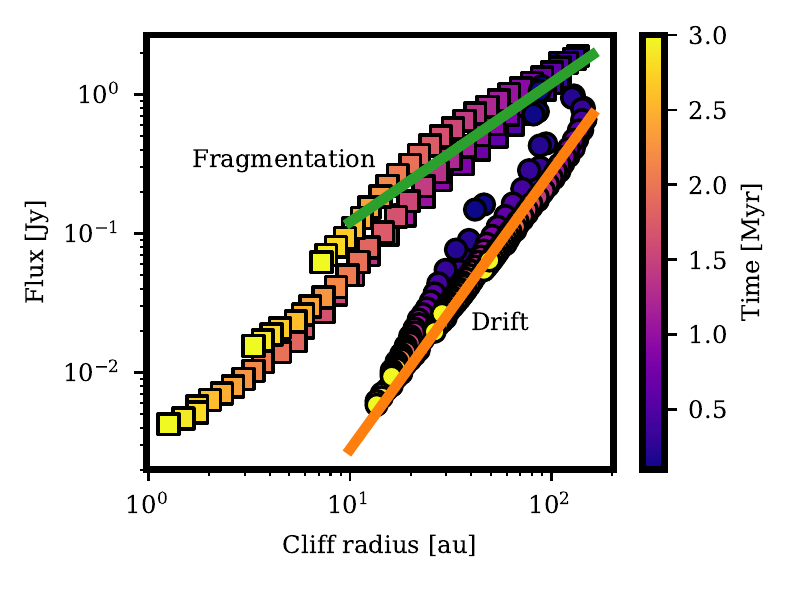}

\vspace{.2cm}

\includegraphics[width=\columnwidth]{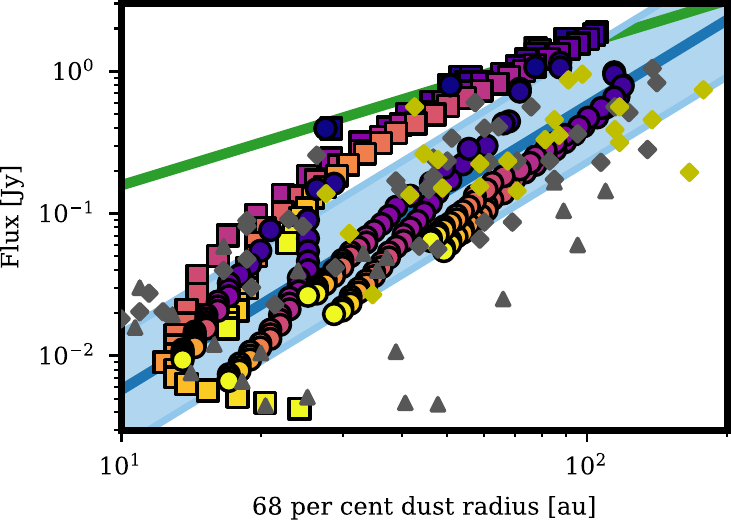}



\caption{{Top:} 850 $\mu$m disc flux for a distance of 140 pc versus opacity cliff radius (see text). Symbols are circles for models in the drift dominated regime ($\alpha = 10^{-3}$ or $10^{-4}$) and squares in the fragmentation dominated regime ($\alpha=0.01$). For reference we show on the plot a quadratic dependence (orange line) and a linear dependence (green line). {Bottom:} as the top panel, but using the 68 per cent flux radius. The blue line shows the observational relation from \citet{Tripathi} and \citet{Andrews2018}, with the associated scatter as the shaded blue region. We plot their data as triangles for stars fainter than 0.2 $L_\odot$ and diamonds for the brighter ones. {The yellow diamonds are the discs showing sub-structure in the DSHARP sample \citep{2018ApJ...869L..41A} and in the Taurus survey \citep{Long2018}}.}
\label{fig:flux_size}
\end{figure}

This expectation is borne out by the full results of our models, where we do not assume optically thin emission nor the Rayleigh-Jeans limit. The top panel of Figure \ref{fig:flux_size} shows at different times the disc flux (assuming a distance of 140 pc) versus the cliff radius. The circles denote the models with low viscosities ($\alpha=10^{-3}$ and $10^{-4}$) and different initial radii (10, 30 and 80 au). The orange line is a quadratic dependence and all the points lie within a factor of two in flux of the same line regardless of the disc parameters, showing that there is excellent agreement with the argument explained above.



\subsubsection{Fragmentation dominated regime}
\label{sec:frag}

In this case it is not straightforward to find a relation between the radius and the flux since the grain size is set by the \textit{gas} surface density rather than that of the \textit{dust}. We can however make the simplifying assumption that the dust-to-gas ratio does not evolve significantly in time. This is reasonable in the fragmentation dominated regime because the grains are smaller and radial drift is less efficient in depleting the disc. With this assumption, using equation \ref{eq:a_frag} we find
\begin{equation}
a_\mathrm{frag} \propto \frac{\Siggas}{ c_s^2} = \frac{\Sigdust}{\epsilon c_s^2} \propto \Sigdust R_\mathrm{cliff}^{1/2}
\end{equation}
and proceeding as before
\begin{equation}
F_\nu \propto \Sigma_{d,\mathrm{cliff}} R_\mathrm{cliff}^{2} T \propto R_\mathrm{cliff}.
\end{equation}
The top panel of figure \ref{fig:flux_size} shows with the squares the evolution of discs with $\alpha=10^{-2}$ in the fragmentation dominated regime. The green line is a linear dependence; the scaling relation we have presented is broadly correct, although it does not fully account for the evolution. In addition, in this case the points do not all lie on the same line, because the flux at a given radius depends on the absolute value of the dust-to-gas ratio. Note also that here we do not take into account that the fragmentation velocity may depend on the grain size \citep{2012A&A...540A..73W}, but this does not change our arguments because the opacity cliff is always met at the same maximum grain size (though it will change the correlation normalisation). 


\subsection{Comparison in observational space}

Observationally, we do not expect to be able to measure exactly the disc cliff radius unless data at high resolution and sensitivity are available. For this reason, in the bottom panel of figure \ref{fig:flux_size} we repeat the analysis using the 68 per cent flux radius. The blue solid line shows the best-fit relation to the observations of \citet{Tripathi}. Models in the drift dominated regime recover correctly both the observed slope and normalisation.

The fragmentation dominated models instead {predict a higher flux for the same disc radius, because they retain more dust mass. While this could be partially mitigated by changing the parameters of the model (see discussion in section \ref{sec:opacity_mass}), a more fundamental issue is that they do not predict, as suggested by our theoretical arguments, a quadratic correlation; the points do not even lie on a single power-law. There is also another important difference}: the 68 per cent flux radius at late times is significantly bigger than the cliff radius. This is because a large fraction of the disc dust mass is in small grains and there is significant flux coming from \textit{outside} the opacity cliff.

While our models correctly capture the overall trend, the observations show a larger scatter than in our models. Assuming that the observed scatter is intrinsic (as will be verified by future, deeper surveys than those currently available), it is possible that some discs are indeed in the fragmentation dominated regime, even if the bulk of the disc population is drift dominated. This would explain discs with average surface brightness that is too \textit{high} for our models. For what concerns the discs with a \textit{low} average surface brightness, we note that their host stars are fainter than 0.2 $L_\odot$ (grey triangles), possibly signalling a change in regime at late spectral types ({maybe because these discs are colder}). Our models show that the correlation is much tighter when using the cliff radius - therefore, ultimately the answer to whether discs are truly in the drift dominated regime will come from high-resolution observations. Here an important prediction of the models (see \autoref{Eqn:FluxR}) is that the discs should have a similar surface brightness ($\sim 0.05$ Jy/arcsec$^2$) at the opacity cliff, i.e. where the surface brightness drops (the ``disc outer edge'').

{Finally, in \autoref{fig:flux_size} we have marked with yellow diamonds discs with known substructures, while for simplicity here we have considered smooth discs. The figure shows the existing selection biases towards bright and large discs. The fact that discs with resolved substructures lie on a correlation derived for smooth discs suggests that substructures may not play a role in shaping the correlation. We will investigate the precise effect of sub-structure in future works.}
  

Summarising, we have provided an explanation as to why models of grain growth in the drift dominated regime predict (see also \citealt{Tripathi}) a quadratic dependence between disc flux and radius. This slope and the normalisation of the correlation are compatible with the observations. We have also highlighted how in this regime the models predict little scatter around the correlation, which is partially in tension with the moderate scatter in the data.



\section{Observational consequences}
\label{sec:consequences}

\subsection{Opacity and mass determination}
\label{sec:opacity_mass}

\begin{figure}
\includegraphics[width=\columnwidth]{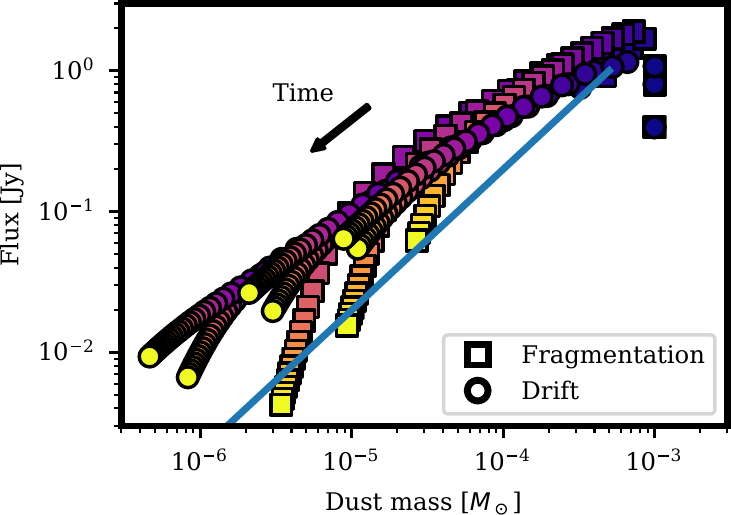}
\caption{Disc flux for a distance of 140 pc versus dust mass in our models. Colours are as in figure \ref{fig:flux_size}. The blue line shows a commonly assumed linear relation between disc flux and mass \citep[e.g.,][]{BeckwithSargent1990}.}
\label{fig:flux_mass}
\end{figure}

Since in our models the disc extends beyond the cliff radius, the sub-mm emission does not trace the full inventory of solid materials, but only the solids with a significant sub-mm opacity. Given that sub-mm fluxes are often used to estimate disc dust masses, this could mean that those masses are \textit{underestimated}. For this reason, we plot in figure \ref{fig:flux_mass} the disc fluxes versus dust masses in our models.

The plot also shows an often employed linear relation between flux and mass, derived by \citet{BeckwithSargent1990} in the optically thin limit using a constant opacity {and a temperature of 20 K}. It is worth noting that {with our assumptions} the opacity at the cliff is one order of magnitude higher than the commonly assumed value of $\sim$ 3 cm$^2$/g. In fact, the higher opacity is more important than the fact that the outer part of the disc is ``invisible'', so that in most of our models the standard assumption \textit{overestimates} the dust mass.

The mass overestimate becomes more pronounced with time, since the dust mass decreases faster than the flux. This is because low mass discs, especially in the drift dominated regime, are small and the emission comes from a hotter part of the disc. On the contrary, in the fragmentation dominated regime the flux eventually drops because the grains become small all over the disc, as we highlighted in section \ref{sec:frag}, and their opacity decreases.

We stress that our models \textit{require} a large opacity to be compatible with the observed flux radius correlation (see \autoref{fig:flux_size}). {In the evolutionary scenario we present in this letter, such high values of the opacity are therefore the only possible choice to make the models compatible with the observations. The opacities we assume are plausible}, but they depend on the (unfortunately unknown) dust composition. If these opacities are correct, the immediate consequence is that the commonly derived dust masses would then be overestimated. The over-estimation makes even more severe the mass budget problem for planet formation \citep{Manara2018}, but possibly reconciles the observed disc fluxes with the significant mass loss due to radial drift. For example, in Taurus the typical dust disc-to-star mass ratio is a few $10^{-5}$ \citep{Andrews2013}, i.e. the typical 850 $\mu$m flux of discs around solar mass stars is tens of mJy; our models naturally explain these values after Myrs of evolution. Finally, note that in these models there is no significant mass in optically thick regions of the disc at small radii.



\subsection{The flux-radius correlation at longer wavelengths}
\label{sec:longer_wavs}

In this section we show that grain growth models in the drift dominated regime predict a flux-radius correlation also at longer wavelengths. As mentioned in section \ref{sec:drift}, our arguments applies as long as there is an opacity cliff and there are grains large enough to be beyond the cliff (i.e., with a size comparable to wavelength). We confirm that qualitatively this is the case also at longer wavelengths (e.g., 3mm; see \autoref{fig:opacity}).

The models also predict how the normalisation should change with frequency. Revisiting \autoref{Eqn:SigD} and \ref{Eqn:FluxR}:
\begin{equation}
F_\nu \approx \upi B_\nu(T) \Sigma_{\rm d,cliff} \kappa_\nu R_{\rm cliff}^2 \propto 
\kappa_\nu a_{\rm cliff} R_\mathrm{cliff}^{2} \nu^2.
\end{equation}
Since the opacity cliff is located where the maximum grain size is a fixed fraction of the observing wavelength, $a_{\rm cliff} \propto \lambda$. Geometric arguments show that the opacity depends on the area-to-mass ratio of the grains, which suggests the opacity at the cliff radius scales as $a_{\rm cliff}^{-1}$. This scaling is upheld by our detailed opacity calculations and can be deduced from \autoref{fig:opacity}. Thus we find that
\begin{equation}
F_\nu / R_{\rm cliff}^2 \propto \nu^2.
\label{eq:fnu_r2_freq}
\end{equation}
Remarkably this quadratic scaling with frequency is the same expected for optically thick emission, {but the two scenarios can be easily distinguished since in the optically thick scenario the disc radius is not a function of frequency. Note that this prediction concerning the frequency dependence of the normalisation of the flux-radius relation is generic to models in which the maximum grain size is set by radial drift and can be tested through ensembles of disc radius and flux measurements at two frequencies. The spectral index of a particular disc (which measures where each disc is located along the flux radius relation at each frequency) is {\it not} a generic prediction of the model as it also depends on the steepness of the dust surface density profile. We thus do not explore spectral index predictions further in this paper, although we note that our models will not differ in this respect from previous studies \citep{Birnstiel2010} where the predicted spectral indices are in tension with the values derived from spatially unresolved multi-frequency data. It is well known that this problem can be mitigated \citep{Pinilla2012} by postulating that discs have sub-structure, as now often (but not always; \citealt{Long2018} find structure only in 30 per cent of the sample) observed \citep{2018ApJ...869L..41A}. The resolution of this discrepancy needs to be further explored through future surveys providing spatially resolved spectral index profiles.}

\section{Conclusions}
\label{sec:conclusions}


In this Letter we have used models of dust evolution to investigate the origin of the recently reported observed correlation between disc flux and radius. Our conclusions are as follows:
\begin{itemize}
    \item {While finite observational sensitivity produces a spurious flux-radius correlation with the observed slope, the observed normalisation is too high to be explained as an observational effect}.
    \item The observed correlation emerges naturally in the drift dominated regime, with the correct slope and normalisation; the latter does not depend on individual disc properties. Conversely, in the fragmentation dominated case the normalisation depends on the dust-to-gas ratio. Therefore, the observed correlation is a sensitive probe of grain growth processes, and suggests that discs are in the drift dominated regime. The model predicts that little scatter around the correlation, which is partially in tension with the observations.
    \item A consequence of being in the drift dominated regime is that the viscosity in discs is relatively low ($\alpha=10^{-3} - 10^{-4}$). This is consistent with studies that attempt to directly measure the level of turbulence in the disc outer parts \citep{2018ApJ...856..117F}.
    \item Explaining the observed disc flux-radius correlation requires a significantly higher opacity than commonly assumed. While plausible, this depends on the unknown dust composition. The observationally derived disc solid masses would then be overestimated.
    \item If discs are in the drift dominated regime, we predict that the correlation is present also at longer wavelengths (e.g., 3mm) and that the normalisation factor scales as the square of observing frequency. This prediction is the same as if discs are optically thick.
\end{itemize}

\section*{Acknowledgements}
This work has been supported by the DISCSIM project, grant agreement 341137 funded by the European Research Council under ERC-2013-ADG. This work was performed in part at Aspen Center for Physics, which is supported by National Science Foundation grant PHY-1607611. This work was partially supported by a grant from the Simons Foundation, from the European Union’s Horizon 2020 research and innovation programme under the Marie Skłodowska-Curie grant agreement No 823823 (DUSTBUSTERS) and by the Munich Institute for Astro- and Particle Physics (MIAPP) of the DFG cluster of excellence "Origin and Structure of the Universe". GR acknowledges support from the Netherlands Organisation for Scientific Research (NWO, program number 016.Veni.192.233). MT has been partially supported by the UK Science and Technology research Council (STFC). GL acknowledges support by the project PRIN-INAF 2016 The Cradle of Life - GENESIS-SKA (General Conditions in Early Planetary Systems for the rise of life with SKA).



\bibliographystyle{mnras}
\bibliography{rfluxcorr}

\begin{thebibliography}{}
\makeatletter
\relax
\def\mn@urlcharsother{\let\do\@makeother \do\$\do\&\do\#\do\^\do\_\do\%\do\~}
\def\mn@doi{\begingroup\mn@urlcharsother \@ifnextchar [ {\mn@doi@}
  {\mn@doi@[]}}
\def\mn@doi@[#1]#2{\def\@tempa{#1}\ifx\@tempa\@empty \href
  {http://dx.doi.org/#2} {doi:#2}\else \href {http://dx.doi.org/#2} {#1}\fi
  \endgroup}
\def\mn@eprint#1#2{\mn@eprint@#1:#2::\@nil}
\def\mn@eprint@arXiv#1{\href {http://arxiv.org/abs/#1} {{\tt arXiv:#1}}}
\def\mn@eprint@dblp#1{\href {http://dblp.uni-trier.de/rec/bibtex/#1.xml}
  {dblp:#1}}
\def\mn@eprint@#1:#2:#3:#4\@nil{\def\@tempa {#1}\def\@tempb {#2}\def\@tempc
  {#3}\ifx \@tempc \@empty \let \@tempc \@tempb \let \@tempb \@tempa \fi \ifx
  \@tempb \@empty \def\@tempb {arXiv}\fi \@ifundefined
  {mn@eprint@\@tempb}{\@tempb:\@tempc}{\expandafter \expandafter \csname
  mn@eprint@\@tempb\endcsname \expandafter{\@tempc}}}

\bibitem[\protect\citeauthoryear{{Andrews}, {Rosenfeld}, {Kraus}  \&
  {Wilner}}{{Andrews} et~al.}{2013}]{Andrews2013}
{Andrews} S.~M.,  {Rosenfeld} K.~A.,  {Kraus} A.~L.,   {Wilner} D.~J.,  2013,
  \mn@doi [\apj] {10.1088/0004-637X/771/2/129}, \href
  {http://adsabs.harvard.edu/abs/2013ApJ...771..129A} {771, 129}

\bibitem[\protect\citeauthoryear{{Andrews}, {Terrell}, {Tripathi}, {Ansdell},
  {Williams}  \& {Wilner}}{{Andrews} et~al.}{2018a}]{Andrews2018}
{Andrews} S.~M.,  {Terrell} M.,  {Tripathi} A.,  {Ansdell} M.,  {Williams}
  J.~P.,   {Wilner} D.~J.,  2018a, \mn@doi [\apj] {10.3847/1538-4357/aadd9f},
  \href {http://adsabs.harvard.edu/abs/2018ApJ...865..157A} {865, 157}

\bibitem[\protect\citeauthoryear{{Andrews} et~al.,}{{Andrews}
  et~al.}{2018b}]{2018ApJ...869L..41A}
{Andrews} S.~M.,  et~al., 2018b, \mn@doi [\apj] {10.3847/2041-8213/aaf741},
  \href {https://ui.adsabs.harvard.edu/\#abs/2018ApJ...869L..41A} {869, L41}

\bibitem[\protect\citeauthoryear{{Ansdell} et~al.,}{{Ansdell}
  et~al.}{2016}]{AnsdellLupusI}
{Ansdell} M.,  et~al., 2016, \mn@doi [\apj] {10.3847/0004-637X/828/1/46}, \href
  {https://ui.adsabs.harvard.edu/#abs/2016ApJ...828...46A} {828, 46}

\bibitem[\protect\citeauthoryear{{Barenfeld}, {Carpenter}, {Sargent}, {Isella}
  \& {Ricci}}{{Barenfeld} et~al.}{2017}]{Barenfeld2017}
{Barenfeld} S.~A.,  {Carpenter} J.~M.,  {Sargent} A.~I.,  {Isella} A.,
  {Ricci} L.,  2017, \mn@doi [\apj] {10.3847/1538-4357/aa989d}, \href
  {https://ui.adsabs.harvard.edu/#abs/2017ApJ...851...85B} {851, 85}

\bibitem[\protect\citeauthoryear{{Beckwith}, {Sargent}, {Chini}  \&
  {Gusten}}{{Beckwith} et~al.}{1990}]{BeckwithSargent1990}
{Beckwith} S. V.~W.,  {Sargent} A.~I.,  {Chini} R.~S.,   {Gusten} R.,  1990,
  \mn@doi [\aj] {10.1086/115385}, \href
  {https://ui.adsabs.harvard.edu/#abs/1990AJ.....99..924B} {99, 924}

\bibitem[\protect\citeauthoryear{{Birnstiel} et~al.,}{{Birnstiel}
  et~al.}{2010}]{Birnstiel2010}
{Birnstiel} T.,  et~al., 2010, \mn@doi [\aap] {10.1051/0004-6361/201014893},
  \href {https://ui.adsabs.harvard.edu/#abs/2010A&A...516L..14B} {516, L14}

\bibitem[\protect\citeauthoryear{{Birnstiel}, {Klahr}  \&
  {Ercolano}}{{Birnstiel} et~al.}{2012}]{Birnstiel2012}
{Birnstiel} T.,  {Klahr} H.,   {Ercolano} B.,  2012, \mn@doi [\aap]
  {10.1051/0004-6361/201118136}, \href
  {https://ui.adsabs.harvard.edu/#abs/2012A&A...539A.148B} {539, A148}

\bibitem[\protect\citeauthoryear{{Booth}, {Clarke}, {Madhusudhan}  \&
  {Ilee}}{{Booth} et~al.}{2017}]{Booth2017}
{Booth} R.~A.,  {Clarke} C.~J.,  {Madhusudhan} N.,   {Ilee} J.~D.,  2017,
  \mn@doi [\mnras] {10.1093/mnras/stx1103}, \href
  {https://ui.adsabs.harvard.edu/#abs/2017MNRAS.469.3994B} {469, 3994}

\bibitem[\protect\citeauthoryear{{Booth}, {Meru}, {Lee}  \& {Clarke}}{{Booth}
  et~al.}{2018}]{Booth2018}
{Booth} R.~A.,  {Meru} F.,  {Lee} M.~H.,   {Clarke} C.~J.,  2018, \mn@doi
  [\mnras] {10.1093/mnras/stx3084}, \href
  {https://ui.adsabs.harvard.edu/#abs/2018MNRAS.475..167B} {475, 167}

\bibitem[\protect\citeauthoryear{{Brauer}, {Dullemond}  \& {Henning}}{{Brauer}
  et~al.}{2008}]{Brauer2008}
{Brauer} F.,  {Dullemond} C.~P.,   {Henning} T.,  2008, \mn@doi [\aap]
  {10.1051/0004-6361:20077759}, \href
  {https://ui.adsabs.harvard.edu/#abs/2008A&A...480..859B} {480, 859}

\bibitem[\protect\citeauthoryear{{D'Alessio}, {Cant{\"o}}, {Calvet}  \&
  {Lizano}}{{D'Alessio} et~al.}{1998}]{Dalessio1998}
{D'Alessio} P.,  {Cant{\"o}} J.,  {Calvet} N.,   {Lizano} S.,  1998, \mn@doi
  [\apj] {10.1086/305702}, \href
  {https://ui.adsabs.harvard.edu/#abs/1998ApJ...500..411D} {500, 411}

\bibitem[\protect\citeauthoryear{{Flaherty}, {Hughes}, {Teague}, {Simon},
  {Andrews}  \& {Wilner}}{{Flaherty} et~al.}{2018}]{2018ApJ...856..117F}
{Flaherty} K.~M.,  {Hughes} A.~M.,  {Teague} R.,  {Simon} J.~B.,  {Andrews}
  S.~M.,   {Wilner} D.~J.,  2018, \mn@doi [\apj] {10.3847/1538-4357/aab615},
  \href {https://ui.adsabs.harvard.edu/#abs/2018ApJ...856..117F} {856, 117}

\bibitem[\protect\citeauthoryear{{Kataoka}, {Okuzumi}, {Tanaka}  \&
  {Nomura}}{{Kataoka} et~al.}{2014}]{Kataoka:2014aa}
{Kataoka} A.,  {Okuzumi} S.,  {Tanaka} H.,   {Nomura} H.,  2014, \mn@doi [\aap]
  {10.1051/0004-6361/201323199}, \href
  {http://adsabs.harvard.edu/abs/2014A%26A...568A..42K} {568, A42}

\bibitem[\protect\citeauthoryear{{Lodato}, {Scardoni}, {Manara}  \&
  {Testi}}{{Lodato} et~al.}{2017}]{Lodato2017}
{Lodato} G.,  {Scardoni} C.~E.,  {Manara} C.~F.,   {Testi} L.,  2017, \mn@doi
  [\mnras] {10.1093/mnras/stx2273}, \href
  {https://ui.adsabs.harvard.edu/#abs/2017MNRAS.472.4700L} {472, 4700}

\bibitem[\protect\citeauthoryear{{Long} et~al.,}{{Long}
  et~al.}{2018}]{Long2018}
{Long} F.,  et~al., 2018, preprint, \href
  {https://ui.adsabs.harvard.edu/#abs/2018arXiv181006044L} {p.
  arXiv:1810.06044} (\mn@eprint {arXiv} {1810.06044})

\bibitem[\protect\citeauthoryear{{Manara} et~al.,}{{Manara}
  et~al.}{2016}]{Manara2016}
{Manara} C.~F.,  et~al., 2016, \mn@doi [\aap] {10.1051/0004-6361/201628549},
  \href {https://ui.adsabs.harvard.edu/#abs/2016A&A...591L...3M} {591, L3}

\bibitem[\protect\citeauthoryear{{Manara}, {Morbidelli}  \& {Guillot}}{{Manara}
  et~al.}{2018}]{Manara2018}
{Manara} C.~F.,  {Morbidelli} A.,   {Guillot} T.,  2018, \mn@doi [\aap]
  {10.1051/0004-6361/201834076}, \href
  {https://ui.adsabs.harvard.edu/#abs/2018A&A...618L...3M} {618, L3}

\bibitem[\protect\citeauthoryear{{Mordasini}, {Molli{\`e}re}, {Dittkrist},
  {Jin}  \& {Alibert}}{{Mordasini} et~al.}{2015}]{MordasiniReview}
{Mordasini} C.,  {Molli{\`e}re} P.,  {Dittkrist} K.~M.,  {Jin} S.,   {Alibert}
  Y.,  2015, \mn@doi [International Journal of Astrobiology]
  {10.1017/S1473550414000263}, \href
  {https://ui.adsabs.harvard.edu/#abs/2015IJAsB..14..201M} {14, 201}

\bibitem[\protect\citeauthoryear{{Mulders}, {Pascucci}, {Manara}, {Testi},
  {Herczeg}, {Henning}, {Mohanty}  \& {Lodato}}{{Mulders}
  et~al.}{2017}]{Mulders2017}
{Mulders} G.~D.,  {Pascucci} I.,  {Manara} C.~F.,  {Testi} L.,  {Herczeg}
  G.~J.,  {Henning} T.,  {Mohanty} S.,   {Lodato} G.,  2017, \mn@doi [\apj]
  {10.3847/1538-4357/aa8906}, \href
  {https://ui.adsabs.harvard.edu/#abs/2017ApJ...847...31M} {847, 31}

\bibitem[\protect\citeauthoryear{{Natta} \& {Testi}}{{Natta} \&
  {Testi}}{2004}]{Natta:2004yu}
{Natta} A.,  {Testi} L.,  2004, in {Johnstone} D.,  {Adams} F.~C.,  {Lin}
  D.~N.~C.,  {Neufeeld} D.~A.,   {Ostriker} E.~C.,  eds,  ASP Conference Series
  Vol. 323, Star Formation in the Interstellar Medium: In Honor of David
  Hollenbach. p.~279

\bibitem[\protect\citeauthoryear{{Natta}, {Testi}, {Calvet}, {Henning},
  {Waters}  \& {Wilner}}{{Natta} et~al.}{2007}]{Natta:2007ye}
{Natta} A.,  {Testi} L.,  {Calvet} N.,  {Henning} T.,  {Waters} R.,   {Wilner}
  D.,  2007, Protostars and Planets V, \href
  {http://adsabs.harvard.edu/abs/2007prpl.conf..767N} {pp 767--781}

\bibitem[\protect\citeauthoryear{{Okuzumi}, {Tanaka}, {Takeuchi}  \&
  {Sakagami}}{{Okuzumi} et~al.}{2011}]{Okuzumi2011}
{Okuzumi} S.,  {Tanaka} H.,  {Takeuchi} T.,   {Sakagami} M.-a.,  2011, \mn@doi
  [\apj] {10.1088/0004-637X/731/2/96}, \href
  {https://ui.adsabs.harvard.edu/#abs/2011ApJ...731...96O} {731, 96}

\bibitem[\protect\citeauthoryear{{Pascucci} et~al.,}{{Pascucci}
  et~al.}{2016}]{Pascucci2016}
{Pascucci} I.,  et~al., 2016, \mn@doi [\apj] {10.3847/0004-637X/831/2/125},
  \href {https://ui.adsabs.harvard.edu/#abs/2016ApJ...831..125P} {831, 125}

\bibitem[\protect\citeauthoryear{{Pinilla}, {Benisty}  \&
  {Birnstiel}}{{Pinilla} et~al.}{2012}]{Pinilla2012}
{Pinilla} P.,  {Benisty} M.,   {Birnstiel} T.,  2012, \mn@doi [\aap]
  {10.1051/0004-6361/201219315}, \href
  {https://ui.adsabs.harvard.edu/\#abs/2012A&A...545A..81P} {545, A81}

\bibitem[\protect\citeauthoryear{{Rosotti}, {Clarke}, {Manara}  \&
  {Facchini}}{{Rosotti} et~al.}{2017}]{Rosotti2017}
{Rosotti} G.~P.,  {Clarke} C.~J.,  {Manara} C.~F.,   {Facchini} S.,  2017,
  \mn@doi [\mnras] {10.1093/mnras/stx595}, \href
  {https://ui.adsabs.harvard.edu/#abs/2017MNRAS.468.1631R} {468, 1631}

\bibitem[\protect\citeauthoryear{{Rosotti}, {Tazzari}, {Booth}, {Testi},
  {Lodato}  \& {Clarke}}{{Rosotti} et~al.}{2019}]{Rosotti2018_rdisc}
{Rosotti} G.~P.,  {Tazzari} M.,  {Booth} R.~A.,  {Testi} L.,  {Lodato} G.,
  {Clarke} C.~J.,  2019, \mnras, in press

\bibitem[\protect\citeauthoryear{{Shakura} \& {Sunyaev}}{{Shakura} \&
  {Sunyaev}}{1973}]{ShakuraSunyaev1973}
{Shakura} N.~I.,  {Sunyaev} R.~A.,  1973, \aap, \href
  {https://ui.adsabs.harvard.edu/#abs/1973A&A....24..337S} {500, 33}

\bibitem[\protect\citeauthoryear{{Tazzari} et~al.,}{{Tazzari}
  et~al.}{2016}]{2016A&A...588A..53T}
{Tazzari} M.,  et~al., 2016, \mn@doi [\aap] {10.1051/0004-6361/201527423},
  \href {https://ui.adsabs.harvard.edu/#abs/2016A&A...588A..53T} {588, A53}

\bibitem[\protect\citeauthoryear{{Testi} et~al.,}{{Testi}
  et~al.}{2014}]{TestiPPVI}
{Testi} L.,  et~al., 2014, in {Beuther} H.,  {Klessen} R.~S.,  {Dullemond}
  C.~P.,   {Henning} T.,  eds, Protostars and Planets VI. p.~339

\bibitem[\protect\citeauthoryear{{Tripathi}, {Andrews}, {Birnstiel}  \&
  {Wilner}}{{Tripathi} et~al.}{2017}]{Tripathi}
{Tripathi} A.,  {Andrews} S.~M.,  {Birnstiel} T.,   {Wilner} D.~J.,  2017,
  \mn@doi [\apj] {10.3847/1538-4357/aa7c62}, \href
  {https://ui.adsabs.harvard.edu/#abs/2017ApJ...845...44T} {845, 44}

\bibitem[\protect\citeauthoryear{{Voelk}, {Jones}, {Morfill}  \&
  {Roeser}}{{Voelk} et~al.}{1980}]{Voelk1980}
{Voelk} H.~J.,  {Jones} F.~C.,  {Morfill} G.~E.,   {Roeser} S.,  1980, \aap,
  \href {https://ui.adsabs.harvard.edu/#abs/1980A&A....85..316V} {85, 316}

\bibitem[\protect\citeauthoryear{{Weidenschilling}}{{Weidenschilling}}{1977}]{Weidenschilling1977}
{Weidenschilling} S.~J.,  1977, \mn@doi [\mnras] {10.1093/mnras/180.1.57},
  \href {https://ui.adsabs.harvard.edu/#abs/1977MNRAS.180...57W} {180, 57}

\bibitem[\protect\citeauthoryear{{Windmark}, {Birnstiel}, {G{\"u}ttler},
  {Blum}, {Dullemond}  \& {Henning}}{{Windmark}
  et~al.}{2012}]{2012A&A...540A..73W}
{Windmark} F.,  {Birnstiel} T.,  {G{\"u}ttler} C.,  {Blum} J.,  {Dullemond}
  C.~P.,   {Henning} T.,  2012, \mn@doi [\aap] {10.1051/0004-6361/201118475},
  \href {https://ui.adsabs.harvard.edu/#abs/2012A&A...540A..73W} {540, A73}

\bibitem[\protect\citeauthoryear{{Zsom}, {Ormel}, {G{\"u}ttler}, {Blum}  \&
  {Dullemond}}{{Zsom} et~al.}{2010}]{Zsom2010}
{Zsom} A.,  {Ormel} C.~W.,  {G{\"u}ttler} C.,  {Blum} J.,   {Dullemond} C.~P.,
  2010, \mn@doi [\aap] {10.1051/0004-6361/200912976}, \href
  {https://ui.adsabs.harvard.edu/#abs/2010A&A...513A..57Z} {513, A57}

\makeatother
\end{thebibliography}



\appendix
\section{An example of finite surface brightness sensitivity}
\label{sec:sensitivity_cut}

Suppose that discs have a power-law surface brightness $I_\nu = A r^{-p}$, where $A$ is some normalisation constant. Given a sensitivity $I_{\nu,cut}$, the disc can be detected up to the radius $r_\mathrm{cut}$ where $A r_\mathrm{cut}^{-p}=I_{\nu,cut}$. The total disc flux $F_\nu$ is given by $\int_0^{r_\mathrm{cut}} 2 \pi I_\nu (r') r' \mathrm{d} r' = 2 \pi I_{\nu,cut} r_\mathrm{cut}^2/(2-p)$, i.e. a quadratic relation between flux and radius. Following \citet{Tripathi}, the relation can be recast in terms of the radius of the disc $r_x$ enclosing a fraction $x$ of the total disc flux, noting that $r_x=x^{1/(p-2)} r_\mathrm{cut}$. This leads to the relation
\begin{equation}
    F_x=\frac{2 \pi }{2-p} x^{p/(2-p)} I_{\nu,cut} r_x^2.
\end{equation}
Therefore the average surface brightness within $r_x$ is $<I_\nu>=2/(2-p) x^{p/(2-p)} I_{\nu,cut}$, i.e. (besides a factor of order unity) the sensitivity threshold. Note that the average surface brightness does not depend on the normalisation constant $A$ and depends only weakly on the exponent $p$, so that a correlation is still expected even if these two parameters vary from disc to disc.

\label{lastpage}
\end{document}